\documentclass[a4paper,11pt]{article}
\usepackage{pos}

\title{Supernova and solar neutrino searches at DUNE}

\author*[a]{C. Cuesta}
\author{the DUNE collaboration }

\affiliation[a]{Centro de Investigaciones Energéticas, Medioambientales y Tecnológicas,
 CIEMAT, \\ 28040, Madrid, Spain}

\emailAdd{clara.cuesta@ciemat.es}

\abstract{The Deep Underground Neutrino Experiment (DUNE) is a next-generation long-baseline experiment exploiting the liquid argon TPC technology. DUNE will have sensitivity to low energy physics searches, such as the detection of supernova and solar neutrinos. DUNE will consist of four modules of 70-kton liquid argon mass in total, placed 1.5 km underground at the Sanford Underground Research Facility in the USA. These modules are being designed considering the specific requirements of the low energy physics searches. As a result, DUNE will have a unique sensitivity for the detection of electron neutrinos from a core-collapse supernova burst, and solar and diffuse supernova background neutrinos can also be detected.}

\FullConference{XVIII International Conference on Topics in Astroparticle and Underground Physics (TAUP2023)\\
 28.08 - 01.09.2023\\
University of Vienna\\}

\begin{document}
\maketitle

\section{Low energy physics in DUNE far detector}

The Deep Underground Neutrino Experiment (DUNE) is a next-generation long-baseline neutrino oscillation experiment~\cite{tdr_v1}. DUNE aims for a high precision measurement of the neutrino oscillation parameters~\cite{DUNE_LBL}, detection of low-energy astrophysical neutrinos~\cite{DUNE_SNB}, and searches beyond the Standard Model~\cite{DUNE_BSM}. 

DUNE will consist of a high intensity, wide-band, neutrino beam from Fermilab (Illinoins, USA); a highly capable neutrino near detector at Fermilab; and a 40-kt fiducial mass far detector (FD) placed 1300-km away, at the Sanford Underground Research Facility (SURF), South Dakota, USA, based on the liquid argon time projection chamber (LArTPC) technology. 

DUNE is anticipated to begin collecting physics data with Phase I, at the beginning of next decade, an initial experiment configuration consisting of two FD modules and a minimal suite of near detector components, with a 1.2\,MW proton beam. The Phase II upgrades necessary to achieve DUNE's physics goals are: addition of FD modules three and four for a total FD fiducial mass of at least 40\,kt, upgrade of the proton beam power to 2.4\,MW and of the near detector.

Low energy astrophysical neutrinos ($\sim$5-100~MeV) from supernovae and the sun will be detected with DUNE FD. Software tools that enable preliminary physics and astrophysics sensitivity studies needed to determine the FD requirements have been developed.

The first FD module will employ the single-phase horizontal-drift (HD) LArTPC technology, where the drift of 3.5~m is horizontal with wrapped-wire readout including two induction and one charge collection anode planes; and the second module will employ the single-phase vertical-drift (VD) LArTPC technology where the drift is vertical over 6~m. Light from photon detectors provide unique information on the timing in both modules. The technology for the other two FD modules corresponding to DUNE Phase II is still to be decided and possibilities to enhance the low energy physics reach are proposed. 

   \begin{figure}[htp]
    \centering
    \includegraphics[width=0.9\textwidth]{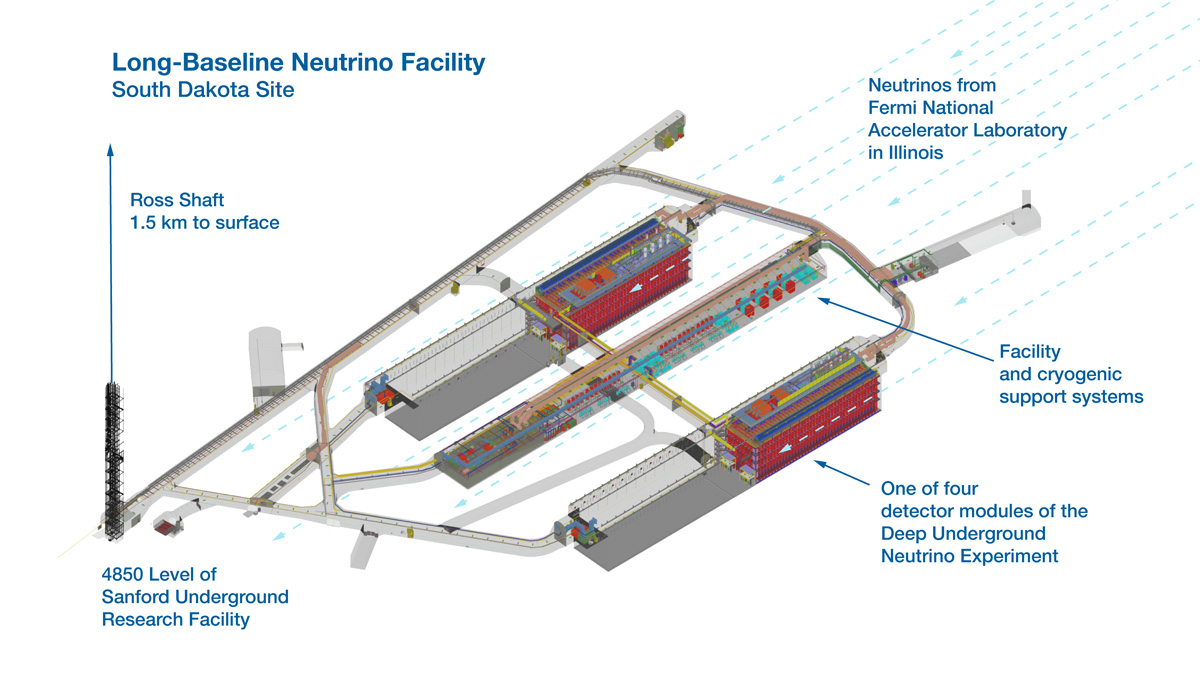}
    \caption{Diagram of the Long-Baseline Neutrino Facility in SURF, South Dakota where the four DUNE Far Detector modules will be placed.}
    \label{fig:1}
    \end{figure}


\section{Supernova neutrino burst detection}

During a core-collapse supernova (SN) explosion, 99\% of the gravitational binding energy of the star  ($\sim$10$^{53}$~ergs) is released as neutrinos and antineutrinos of all flavors. The neutrino signal starts with a short, sharp \textit{neutronization} burst primarily composed of $\nu_e$. This is followed by an \textit{accretion} phase lasting several hundred milliseconds, and then a \textit{cooling} phase which lasts about 10~s and represents the bulk of the signal, roughly equally divided among all flavors of neutrinos and antineutrinos. The flavor content and spectra of neutrinos change throughout these phases, so the supernova’s evolution can be mapped out using the neutrino signal. Information about the progenitor, the collapse, the explosion, and the remnant, as well as information about neutrino properties, are contained in this signal. 

\begin{figure}[htp]
    \centering
    \includegraphics[width=0.5\textwidth]{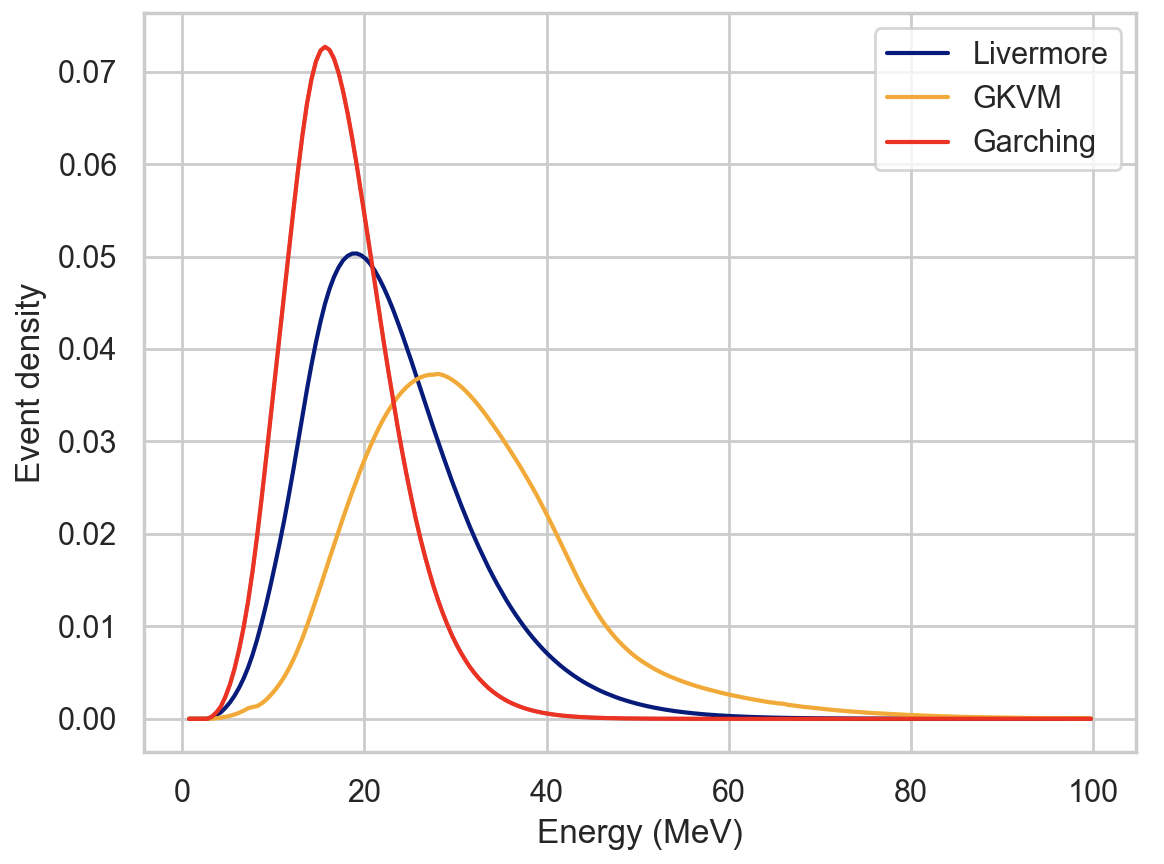}
    \caption{Probability density as a function of neutrino energy for $\nu_e$ charged-current events interacting in DUNE for three SNB models \cite{Totani_1998, Gava:2009pj, Huedepohl:2009wh}.}
    \label{fig:2}
    \end{figure}

Hence, the goal of DUNE is to measure and study the time and energy spectra of the supernova neutrino burst (SNB). The expected $\nu_e$ energy spectrum for three supernova models is shown in Figure~\ref{fig:2} for events scattering in DUNE.

The predicted event rate from a SNB is calculated by folding together expected neutrino differential energy spectra and cross sections for the relevant channels. Details of the simulation and reconstruction are provided in~\cite{DUNE_SNB}. Predicted event rates vary up to an order of magnitude among different SNB models, and rates will scale as the inverse square of supernova distance. As a benchmark, DUNE would observe $\sim$3300 $\nu_e$ charged-current, 210 electron-scattering and 160 $\bar{\nu_e}$ charged-current events for a core collapse supernova at 10\,kpc in the GKVM model~\cite{Gava:2009pj}, assuming 40\,kt fiducial mass of argon for DUNE.  DUNE will be
uniquely sensitive to the electron-neutrino-flavour component of the SNB.

As SN within our own Galaxy are expected to occur once every few decades, it is critical that experiments like DUNE are prepared to capture as much data as possible when one does occur, which drives the requirements for detector livetime, DAQ and triggering systems, and reconstruction capabilities for low-energy events. DUNE’s expected energy resolution is around 10-20\% for energies in the few tens of MeV range. In DUNE, a highly efficient trigger on a SNB within our galaxy can be done using either TPC or photon detection system information exploiting the time coincidence of multiple signals over a timescale matching the SN luminosity evolution.


\section{Solar neutrinos and diffuse supernova neutrino background}

Detection of solar neutrinos in DUNE is challenging in a LArTPC because of relatively high intrinsic detection energy thresholds for the charged-current interaction on argon of about 5\,MeV and because radioactive backgrounds are large and although can be fiducialized, they can affect the triggering capability. DUNE may be capable of performing world-leading measurements of fundamental neutrino properties. Measuring the $^8$B neutrino flux with a 15.0-MeV endpoint, DUNE can measure the solar mass splitting $\Delta m^2_{12}$ and mixing angle $\theta_{12}$ with higher precision than the previous measurements, enabling more stringent tests of the low significance tension between terrestrial and solar measurements of these parameters. DUNE may also be able to provide first measurements of the $hep$ neutrino flux, extending to 18.8~MeV, thanks to its particular sensitivity to $\nu_e$ interactions. Detailed studies of solar neutrino detection capability are underway in DUNE. 

Similarly, DUNE can search for the Diffuse Supernova Neutrino Background (DSNB)~\cite{Beacom:2010kk} at energies just above the endpoint of the solar neutrino spectrum. As DUNE is primarily sensitive to the $\nu_e$ component, DUNE will be the only running experiment with sensitivity to the neutrino component of the DSNB.

\section*{Acknowledgments}

Work produced with the support of a 2023 Leonardo Grant for Researchers in Physics, BBVA Foundation and part of the RYC2021–031667–I founded by MCIN/AEI/10.13039/501100011033 and the European Union NextGenerationEU/PRTR.

\bibliographystyle{JHEP}
\bibliography{skeleton}

\end{document}